\documentclass[letterpaper]{article} % DO NOT CHANGE THIS
\usepackage{aaai2026}  % DO NOT CHANGE THIS
\usepackage{booktabs}
\usepackage{times}  % DO NOT CHANGE THIS
\usepackage{helvet}  % DO NOT CHANGE THIS
\usepackage{courier}  % DO NOT CHANGE THIS
\usepackage[hyphens]{url}  % DO NOT CHANGE THIS
\usepackage{graphicx} % DO NOT CHANGE THIS
\usepackage{natbib}  % DO NOT CHANGE THIS AND DO NOT ADD ANY OPTIONS TO IT
\usepackage{caption} % DO NOT CHANGE THIS AND DO NOT ADD ANY OPTIONS TO IT
\usepackage{algorithm}
\usepackage{algorithmic}
\usepackage{amsmath}

\usepackage{newfloat}
\usepackage{listings}
\DeclareCaptionStyle{ruled}{labelfont=normalfont,labelsep=colon,strut=off} % DO NOT CHANGE THIS
\floatstyle{ruled}
\newfloat{listing}{tb}{lst}{}
\floatname{listing}{Listing}
\title{Masked Autoencoders Learn Perception-Relevant Representations \\ from Resting State Neural Data}
\author{
    Aleksandr Kovalev\equalcontrib,
    Antonio Lozano\equalcontrib,
    Fabrizio Grani\equalcontrib,
    Cristina Soto Sanchez,
    Leili Soo,
    Rocío López-Peco,
    Adrian Villamarin-Ortiz,
    Roberto Morollón Ruiz,
    María del Mar Ayuso Arroyave,
    Alfonso Rodil,
    Eduardo Fernández
}
\affiliations{
    Bioengineering Institute and Cátedra Bidons Egara, University Miguel Hernández\\
    Elche, Spain\\
    akovalev@umh.es
}

\usepackage{bibentry}
\begin{document}

\maketitle

\begin{abstract}

Clinical neuroprosthetics face a data bottleneck: labeled perception trials are scarce while hours of spontaneous neural activity are largely underutilized. Here, we test whether self-supervised learning can use these unlabeled datasets to improve perception decoding. We pretrained a masked autoencoder on 14.6 hours of spontaneous multiunit activity from an intracortical array in a blind participant's V1. The model captured interpretable brain structure without supervision: V1's spatial organization and perceptual state separation both emerged purely from its latent representations.

To test these features, we used linear probing (logistic regression on the frozen latents) to measure performance on the data with stimulation. Perception decoding accuracy reached 84.1\% on a general psychometric task. On the more difficult threshold-level task, accuracy reached 64.0\%. This work shows that spontaneous cortical activity is not noise; it contains rich, task-relevant structure. Unsupervised pretraining on this data is a promising strategy to improve neural decoding.
\end{abstract}

\section{Introduction}
Decoding subjective perception from neural activity requires labeled training data, pairing neural recordings with perceptual reports. In clinical neuroprosthetics, such data are scarce: participant availability is limited, experimental sessions are short, and subjective reports are inherently difficult to measure. Meanwhile, tens to hundreds of hours of spontaneous neural activity accumulate during rest periods and between experimental blocks. Data that are routinely left underutilized. This creates a fundamental imbalance that limits neuroengineers from training decoding models: abundant unlabeled recordings, limited labeled examples. Here, we demonstrate that self-supervised learning in spontaneous V1 activity can improve perception decoding.

This data bottleneck is most severe at perceptual threshold. Visual neuroprosthetics use intracortical microstimulation to evoke phosphenes (perceived flashes of light) in blind individuals, with recent demonstrations of shapes and letters percepts \cite{Fernandez2021Visual, Chen2020Shape}. But perception at threshold is inconsistent for identical stimulation sometimes produces a phosphene, sometimes it does not. This variability is reflected by spontaneous fluctuations in cortical state \cite{vanVugt2018Threshold}. Predicting whether stimulation will be perceived would enable adaptive systems that learn to stimulate during favorable cortical states. Yet this is precisely where supervised decoders fail: with limited labeled data, they cannot distinguish the subtle neural states that determine perception versus non-perception at threshold.

Self-supervised learning offers a potential solution. We hypothesize that spontaneous cortical activity explores the same low-dimensional manifolds that govern stimulus-evoked responses. If true, we can pretrain on abundant unlabeled resting data to learn the structure of cortical state space, then use those representations for perception decoding. Masked autoencoders (MAEs) work by reconstructing randomly masked input, achieving strong results in multiple domains \cite{He2021Masked, Tong2022VideoMAE, Baade2022Masked}, including recent work on neural data \cite{Zhuang2023NDT2, Sarid2023Universal}.

\begin{figure}[t]
\centering
\includegraphics[width=0.45\textwidth]{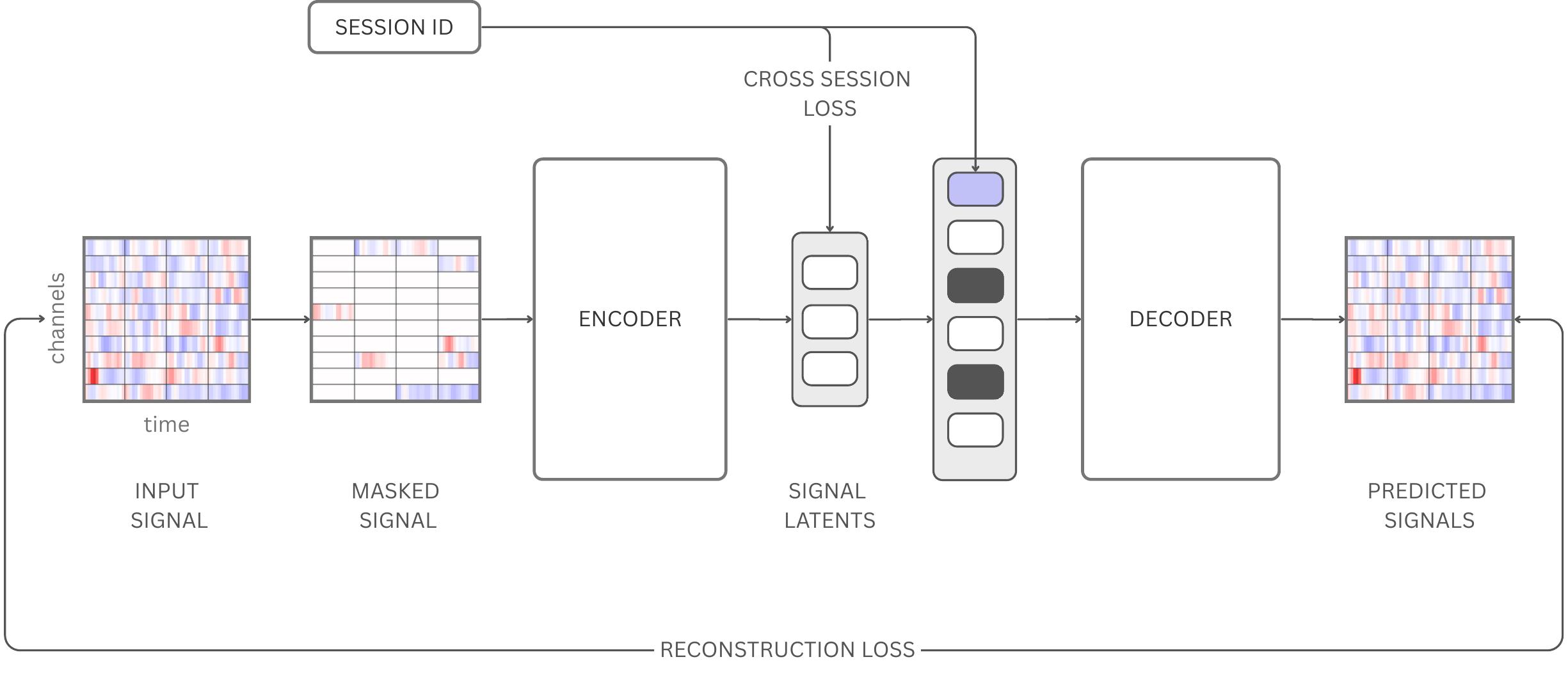}
\caption{\textbf{Self-supervised pretraining architecture.} Input signals are randomly masked. The encoder processes only visible patches. The decoder receives encoder outputs, learned mask tokens, and a session embedding. }
\label{fig:model}
\end{figure}

We pretrained a masked autoencoder on 14.6 hours of spontaneous activity from a 96-channel Utah array implanted in the area V1 of a blind participant. Critically, the model never saw neural responses to microstimulation during pretraining. The learned representations captured interpretable brain organization, spatial layout and perceptual state separation, despite never receiving explicit supervision. 

 We then evaluated these representations on two perception decoding tasks with linear probing. On a general psychometric task, the pretrained features achieved 84.1\% accuracy. More critically, when evaluated on perceptual report trials at the 50\% threshold, where identical stimuli produced variable outcomes, pretraining improved decoding accuracy achieving 64.0\%, demonstrating that spontaneous dynamics contain structure relevant to perception.

\section{Methods}
We recorded from a 96-channel Utah array in V1 of a participant with acquired blindness. Three datasets gave us complementary views of the same neural population: spontaneous activity to learn dynamics, psychometric curves to map perception, and threshold trials where identical stimuli produced variable perception.

\subsection{Datasets}

\subsubsection{Resting State.} 252 sessions of spontaneous activity (14.6 hours total) with no stimulation. We used this data to train our model.
\subsubsection{Psychometric.} Single-electrode microstimulation across 12 electrodes at six current levels (1-81 $\mu$A and 3--5 repetitions) to map how stimulation intensity related to perception.
\subsubsection{Threshold.} Stimulation at perceptual threshold on 4 electrodes where the same current sometimes triggered perception and sometimes did not. For each electrode, we used the current level that produced perception on 50\% of trials. We did 100-300 repetitions in each session.

For both stimulation datasets, we used single-electrode stimulation with 50 biphasic pulses with a pulse width of 170~$\mu$s and interpulse interval of 60~$\mu$s at 300~Hz (167~ms total train duration). We extracted 128 ms of neural activity starting 170 ms after stimulation onset to capture the population response without artifact. We used per session temporal splits (80\% early trials for training, 20\% late for testing) to test generalization over time. Full recording parameters are shown in Table~\ref{tab:datasets}.

\subsection{Data Preprocessing}
We extracted multi-unit activity (MUA) using a standard preprocessing pipeline (Table~\ref{tab:preproc}). The key difference was the normalization strategy, which depended on whether the data contained stimulation artifacts.
For resting state data, we applied continuous z-score normalization with a rolling 10-second window. For stimulation data, we used robust z-score normalization (median and median absolute deviation) because stimulation artifacts would blow up standard mean/std calculations. We linearly interpolated across blanked stimulation periods before bandpass filtering to avoid edge artifacts.

\subsection{Self-Supervised Pretraining}

\subsubsection{Model Architecture}

We use a standard MAE setup with registers and masking. We add three common modifications to the architecture: pre-layer RMSNorm, SwiGLU, and QKNorm - all standard in the field.
For positional encoding, we use separate learnable embeddings for temporal and spatial positions, then sum them for each signal embedding.
During training, we randomly sample signal windows of 96 electrodes × 128ms. We patchify with temporal patch size = 16 and spatial patch size = 1. Following standard MAE design, we randomly mask 75\% of patches. The encoder processes only the visible patches, and the decoder reconstructs the masked ones.
The decoder receives three components: (1) the encoder's output latents from visible patches, (2) learned mask tokens for the masked positions, and (3) a learned session embedding token. The decoder then reconstructs the masked patches from this combined input. Full architectural details are in Figure~\ref{fig:model} and Table~\ref{tab:arch}.
In the following experiments, we obtain the final latent representation by averaging all encoder outputs, producing a 768-dimensional latent vector

\subsubsection{Training Objective}
We combine two loss terms to learn representations that generalize between sessions while preserving  neural dynamics.
First, we used a reconstruction loss with two components: MSE loss and correlation loss between predicted and true signals. We calculate MSE only on masked patches.
Second, we use a cross-session loss implemented with an inverted SigLip formulation. This loss pulls together samples from different sessions and pushes apart samples from the same session. This forces the model to pay less attention to session-specific features and learn representations that capture the underlying neural dynamics instead.

\begin{equation}
\mathcal{L} = \mathcal{L}_{\text{MSE}} + 0.01 \cdot \mathcal{L}_{\text{Cross-session}} + 0.1 \cdot \mathcal{L}_{\text{Corr}}
\end{equation}

\subsubsection{Training Details}
We train the model for 1M steps with batch size 32 using the AdamW optimizer. We set the learning rate to $1 \times 10^{-4}$ with 5k warmup steps, then decay it to $1 \times 10^{-5}$ using a cosine schedule. We select the best checkpoint by evaluating reconstruction loss on a held-out validation set consisting of the last 4 days of recording.

\subsubsection{Evaluation}
For decoding, we froze the pretrained encoder and extracted 768-dimensional features by averaging outputs across patches. We then trained L2-regularized logistic regression to predict perception in the psychometric and threshold level datasets. We compared against two baselines: MUA (12,288-dimensional flattened data) and PCA on MUA signals (768 dimensions, matching encoder output).

\begin{figure*}[t]
\centering
\includegraphics[width=0.95\textwidth]{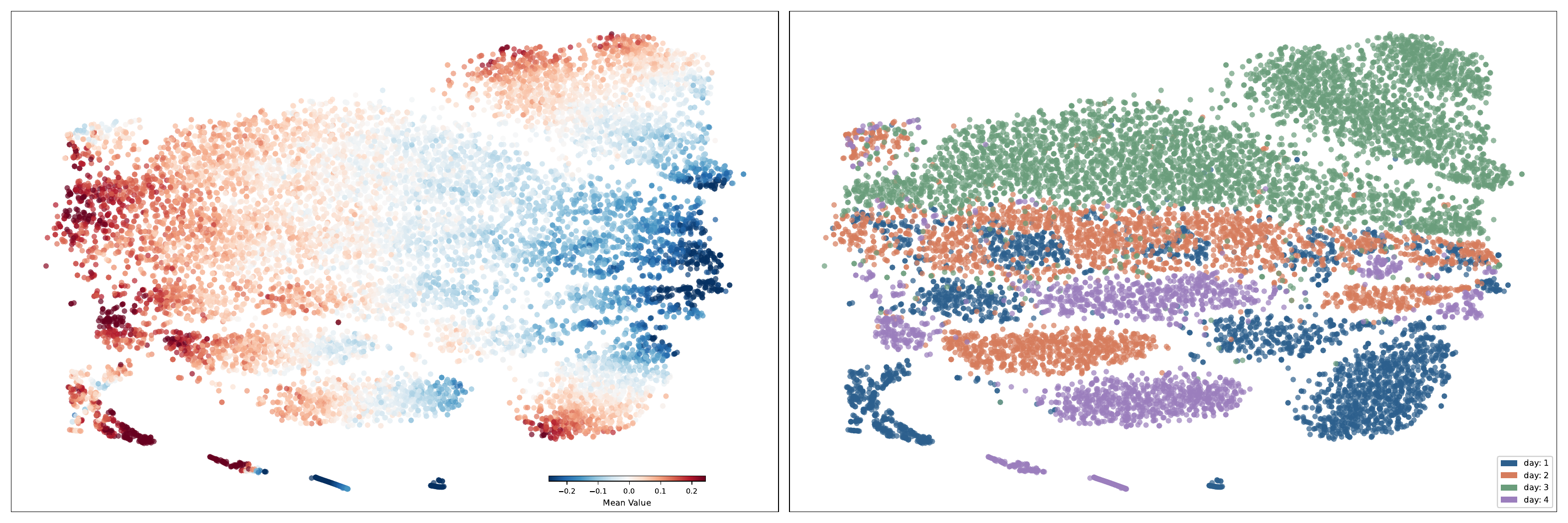} 
\caption{\textbf{t-SNE projection of spontaneous activity features. }(Left) Colored by mean neural activity. (Right) Colored by recording day. }
\label{fig:latent_full_resting}
\end{figure*}

\begin{figure*}[t]
\centering
\includegraphics[width=0.95\textwidth]{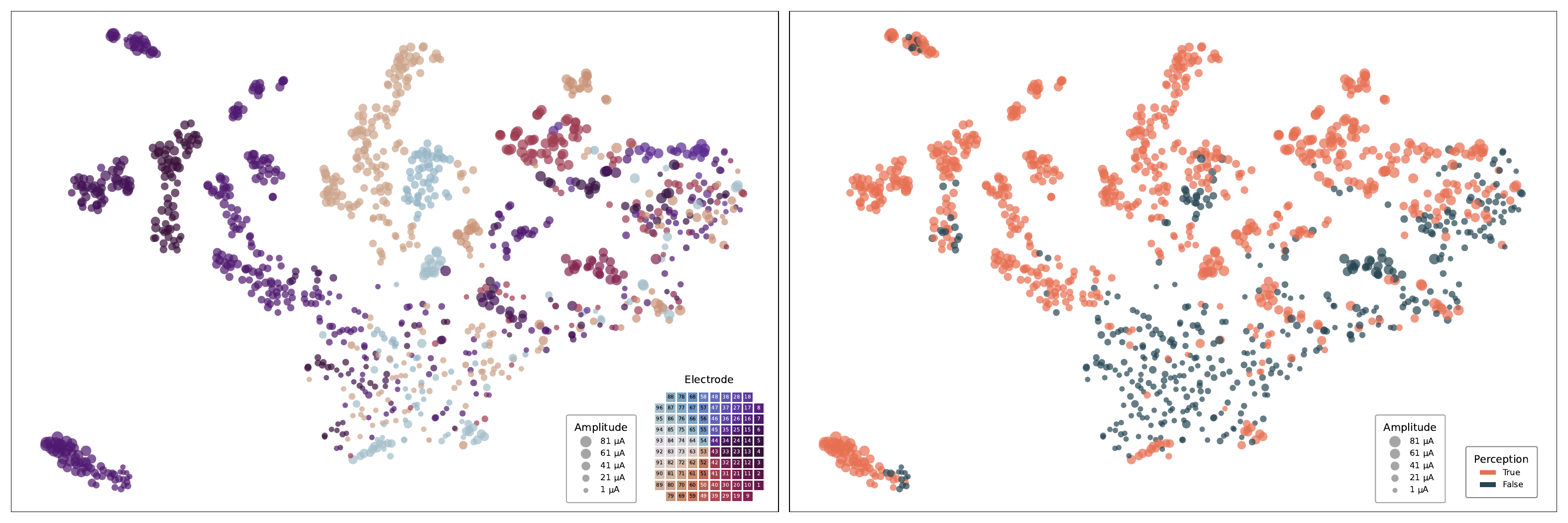} 
\caption{\textbf{Learned representations capture spatial and perceptual structure.} t-SNE projection of MAE features for psychometric trials. (Left) Colored by electrode and stimulation amplitude. Spatially adjacent electrodes cluster together, probably reflecting V1's retinotopic organization. (Right) Colored by perception outcome. Perceived (orange) and missed (dark blue) trials show partial separation, indicating learned features reflect perceptual states despite no explicit supervision during pretraining.}
\label{fig:latent_full}
\end{figure*}

\section{Results}
\subsection{Learned Representations Captured Spatial and Perceptual Structure}

We trained a masked autoencoder on resting state V1 activity. The resulting representations captured complex structure in the spontaneous dynamics; for instance, held-out spontaneous sessions with similar high mean MUA and spatiotemporal patterns were separated into distinct clusters in the latent space.

To understand if these representations learned \textit{only} from spontaneous data were relevant to perception, we passed the \textbf{psychometric dataset} through the frozen, pretrained encoder. We then visualized these resulting latent representations using t-SNE (Figure~\ref{fig:latent_full_resting}). Two patterns emerged, even though the model was never trained on stimulation or perception labels:

\subsubsection{Spatial organization}
When colored by electrode identity (Figure~\ref{fig:latent_full}, left), trials from nearby electrodes clustered together in latent space. The model discovered this organization purely from spontaneous data, which indicated it learned anatomically meaningful features. Additionally, stimulation amplitude (shown by point size) systematically shifted the position within these electrode clusters, suggesting that the learned features also reflected current injection magnitude.

\subsubsection{Perceptual state separation}
When colored by perception outcome (Figure~\ref{fig:latent_full}, right), perceived ('True') and missed ('False') trials occupied partially distinct regions of latent space. Substantial overlap remained, which was consistent with the inherent stochasticity of near-threshold perception.

\subsection{Pretraining on Spontaneous Activity Improved Perception Decoding}

\begin{table}[h]
\centering
\caption{Decoding Performance: Psychometric vs. Threshold-level Perception. Balanced accuracy was used as the metric.}
\label{tab:combined_decoding}
\begin{tabular}{lcc}
\hline
& Psychometric & Threshold-level \\
\hline
Neural data & 80.2\% & 60.0\% \\
Neural data + PCA & 82.4\% & 61.3\% \\
\textbf{MAE (ours)} & \textbf{84.1\%} & \textbf{64.0\%} \\
\hline
\end{tabular}
\end{table}

To test if the model generalized to perception decoding, we trained an L2-regularized logistic regression classifier to predict perception outcomes on top of the features.

We first evaluated on the psychometric dataset. MAE features achieved 84.1\% balanced accuracy, outperforming both MUA neural data (80.2\%) and PCA (82.4\%) (Table~\ref{tab:combined_decoding}). The improvement over PCA indicated the model learned relevant nonlinear structure that linear compression missed.

The critical test used the threshold-level dataset. Here, stimulation amplitude was fixed at the perceptual threshold (50\% detection), and perception varied stochastically across trials. MAE features achieved 64.0\% accuracy, compared to 60.0\% for MUA neural data and 61.3\% for PCA (Table~\ref{tab:combined_decoding}). This result demonstrated that pretraining on spontaneous activity extracted features useful for threshold-level decoding.

\subsection{Analysis of Threshold-Level Representations}
\begin{figure}[t]
\centering
\includegraphics[width=0.45\textwidth]{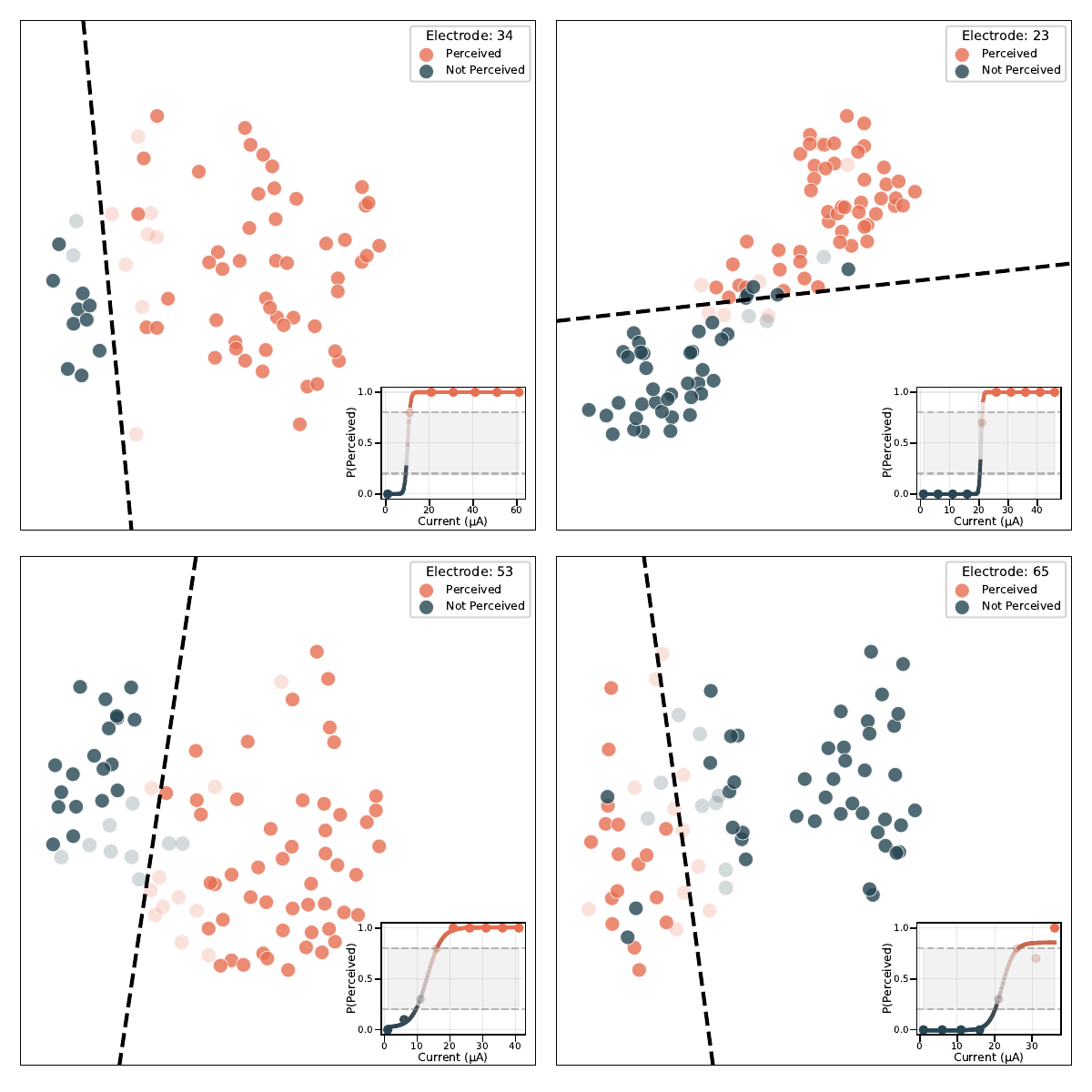}
\caption{\textbf{Per-session t-SNE of stimulation responses showed clean separation} (orange=perceived, blue=missed). Mixing only occurred in the 20-80\% psychometric range where perception itself was uncertain.}
\label{fig:latent_session}
\end{figure}

To understand the difficulty of threshold decoding, we analyzed the latent space. As shown previously (Figure~\ref{fig:latent_full}, right), perceived and missed trials occupied partially distinct regions, but substantial overlap remained. This was consistent with the inherent stochasticity of near-threshold perception.

This variability became clearer in per-session analysis (Figure~\ref{fig:latent_session}). Some electrodes such as Electrode 34, showed a relatively clean separation between perceived and missed trials. Others, like Electrode 65, showed heavy mixing, with decision boundaries cutting through dense, overlapping zones.

This visualization highlighted the core challenge of threshold decoding: even with features that captured meaningful structure, the inherent stochasticity of perception led to uncertainty. Some trials simply could not be classified from the neural state alone, as nearly identical states could lead to different perceptual outcomes. The 64.0\% accuracy we achieved was likely constrained by this neural noise.

\section{Discussion}
We tested if abundant, unlabeled spontaneous neural activity could overcome the data bottleneck in perception decoding. To do this, we trained a masked autoencoder on 14.6 hours of spontaneous V1 activity without any task or perception labels. The approach was effective: the learned representations captured meaningful brain structure, including V1's spatial organization and a separation between "perceived" and "missed" trials. This unsupervised pretraining translated directly to better performance, achieving 84.1\% accuracy on a general psychometric task and 64.0\% on the difficult threshold-level task.

\subsubsection{Limitations and Open Questions} Our main limitation is that these findings are from a single subject. We don't know if this approach will work for other participants, so further replication is needed. Additionally, we're not entirely sure why this pretraining works. We speculate that spontaneous activity explores the same fluctuating brain states, replays some vision experience, or establishes a baseline manifold against which stimulus evoked responses can be contrasted. Also, our current 128ms analysis window might be too brief to capture these slower changes, which is an important area for future work.

\subsubsection{Future Directions} The next direct step is a hybrid training approach. We intentionally used only spontaneous data in this study to isolate its contribution to representation learning. Now, we can combine this resting-state data with unlabeled stimulation-trial data. A model trained on both datasets could learn the structure of spontaneous and task evoked activity, which would likely improve decoding performance further.

\subsubsection{The Core Finding} Our core finding is that spontaneous activity provides a valuable pretraining signal. We showed that a self-supervised model, trained without any labels, could learn fundamental properties of the brain, like its spatial layout and states related to perception. This demonstrates that vast, unlabeled neural datasets can be effectively leveraged with this approach.

\bibliography{refs}

\newpage
\appendix

\section{Supplemental Materials}

\subsection{Methods}

% Requires the 'booktabs' package (which AAAI style files load)
\begin{table*}[t]
\centering
\caption{Dataset characteristics across resting state and stimulation paradigms.}
% l: left-aligned (labels)
% r: right-aligned (numbers)
\begin{tabular}{lrrr}
\toprule
& \textbf{Resting State} & \textbf{Psychometric} & \textbf{Threshold} \\
\midrule
Sessions & 252 & 22 & 19 \\
Unique dates & 96 & 13 & 11 \\
Recording duration & 14.6 hrs & 0.8 hrs & 1.3 hrs \\

Trials & --- & 1,205 & 1,760 \\

Stimulated electrodes & --- & \begin{tabular}[t]{@{}c@{}}6, 14, 20, 23, 25, 34, \\ 42, 44, 53, 65, 70, 96\end{tabular} & 6, 25, 53, 65 \\
\bottomrule
\end{tabular}
\label{tab:datasets}
\end{table*}

\begin{table*}[h]
\centering
\caption{Preprocessing pipeline for multi-unit activity extraction}
\label{tab:preproc}
\begin{tabular}{ll}
\hline
Step & Operation \\
\hline
1 & Common average reference (96 channels) \\
2 & Clip to $\pm 5000~\mu$V \\
3 & Bandpass filter (500--5000~Hz) \\
4 & Full-wave rectification \\
5 & Lowpass filter (200~Hz) \\
6 & Downsample to 1~kHz (average pooling) \\
7 & Normalize (rolling z-score or robust z-score) \\
\hline
\end{tabular}
\end{table*}

\begin{table*}[h]
\centering
\caption{Model architecture}
\label{tab:arch}
\small
\begin{tabular}{lcc}
\hline
Component & Encoder & Decoder \\
\hline
Layers & 24 & 8 \\
Hidden dimension & 768 & 768 \\
Attention heads & 16 & 16 \\
Head dimension & 64 & 64 \\
Masking ratio & 75\% & - \\
Input & Visible patches & All patches + session ID \\
Parameters & 246M & 83M \\
\hline
\end{tabular}
\end{table*}

\end{document}